\documentclass{article}

\PassOptionsToPackage{authoryear, comma, round}{natbib}

\usepackage[preprint]{neurips_2022}

\usepackage[utf8]{inputenc} 
\usepackage[T1]{fontenc}    
\usepackage{hyperref}       
\usepackage{url}            
\usepackage{booktabs}       
\usepackage{amsfonts}       
\usepackage{nicefrac}       
\usepackage{microtype}      
\usepackage{xcolor}         
\usepackage{graphicx}
\usepackage{bm}
\usepackage{amsmath,amssymb}
\usepackage{ulem}

\title{SpeechGen: Unlocking the Generative Power of Speech Language Models with Prompts}

\author{%
  Haibin Wu*, Kai-Wei Chang*, Yuan-Kuei Wu*, Hung-yi Lee\\
  Graduate Institute of Communication Engineering, National Taiwan University \\
}

\begin{document}

\maketitle

\begin{abstract}
Large language models (LLMs) have gained considerable attention for Artificial Intelligence Generated Content (AIGC), particularly with the emergence of ChatGPT. However, the direct adaptation of continuous speech to LLMs that process discrete tokens remains an unsolved challenge, hindering the application of LLMs for speech generation. The advanced speech LMs are in the corner, as that speech signals encapsulate a wealth of information, including speaker and emotion, beyond textual data alone. Prompt tuning has demonstrated notable gains in parameter efficiency and competitive performance on some speech classification tasks. However, the extent to which prompts can effectively elicit generation tasks from speech LMs remains an open question. In this paper, we present pioneering research that explores the application of prompt tuning to stimulate speech LMs for various generation tasks, within a unified framework called SpeechGen, with around 10M trainable parameters. The proposed unified framework holds great promise for efficiency and effectiveness, particularly with the imminent arrival of advanced speech LMs, which will significantly enhance the capabilities of the framework. The code and demos of SpeechGen will be available on the project website: \url{https://ga642381.github.io/SpeechPrompt/speechgen}

\end{abstract}

\begin{figure}[h!]
    \centering
    \includegraphics[width=0.98\columnwidth]{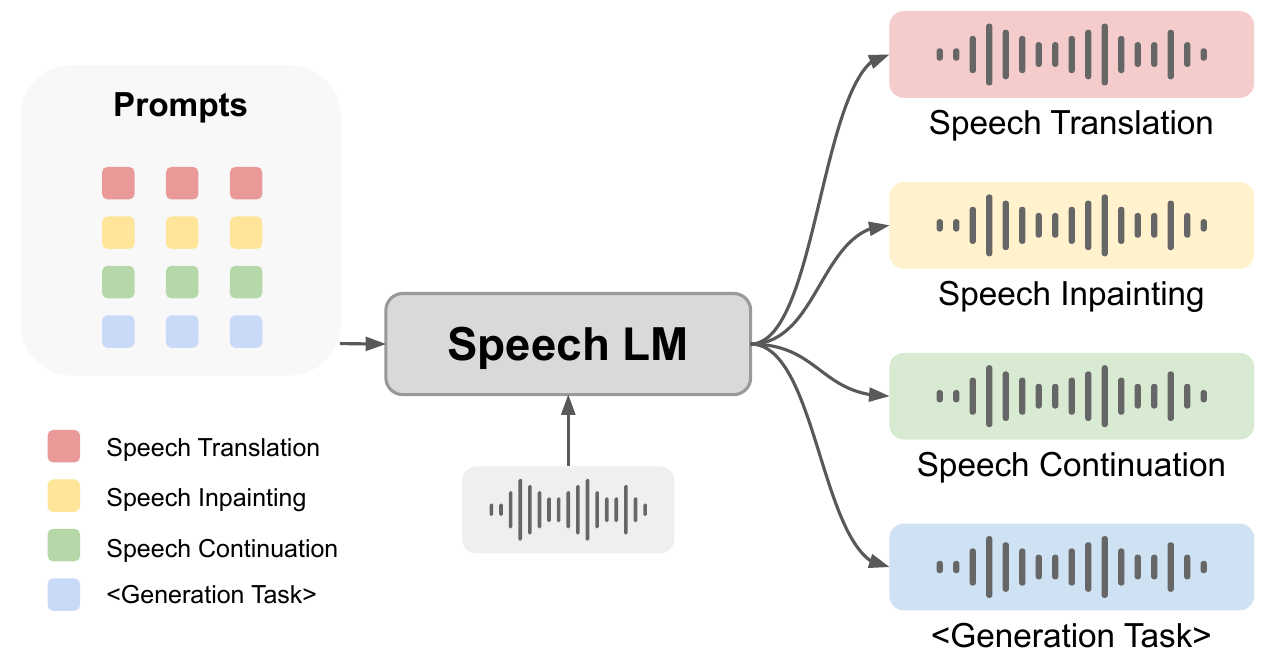}
    \caption{SpeechGen: Prompting Speech Language Model to perform speech generation tasks. When provided with a specific prompt and speech as inputs, the speech LM can perform a specific task.}
\end{figure}

\section{Introduction}
\label{sec:intro}
Large language models (LLMs) have garnered substantial attention for Artificial Intelligence Generated Content (AIGC), particularly with the advent of ChatGPT. 
However, there remains a notable disparity between existing text-based LLMs and the goal of achieving AIGC. 
The challenge lies in the direct adaptation of continuous speech, to LLMs that primarily process discrete tokens. 
This limitation hampers the effective utilization of LLMs for speech generation purposes.

Speech signals encompass a wealth of information, extending beyond the confines of textual data, including details about the speaker and emotional nuances, and para-linguistic information such as prosody. 
Consequently, there is a pressing necessity to design advanced speech language models (speech LMs) capable of capturing and leveraging this rich information. 
While speech LMs are still in their early stages compared to text-based language models, their potential is promising due to the richer information contained in speech data compared to text.

Researchers have been actively investigating the potential of the prompting paradigm \citep{liu2021pre} to harness the capabilities of pre-trained language models. 
Prompting entails the utilization of task-specific templates or instructions to guide a pre-trained LM, while keeping its architecture and parameters unchanged. 
Prompt tuning, a widely adopted technique in natural language processing (NLP), involves fine-tuning a small subset of parameters to steer a pre-trained LM towards specific downstream tasks. 
This technique has gained popularity in NLP due to its effectiveness and efficiency. 
In the realm of speech processing, SpeechPrompt ~\citep{DBLP:conf/interspeech/ChangT0L22} has demonstrated noteworthy enhancements in parameter efficiency and has achieved competitive performance on diverse speech classification tasks.

However, the effectiveness of prompts in eliciting generation tasks from speech LMs remains an open question.
In this paper, we present a unified framework designed to motivate speech LMs for generation tasks. The proposed prompt framework offers several advantages:
\begin{itemize}
    \item \textbf{Textless}: Our framework, as well as the underlying speech language models it relies on, operates independently of text data. The significance of textless speech generation should not be underestimated, considering the time-consuming process of obtaining labeled text-speech pairs, and the absence of text in certain languages. The textless property enables powerful speech generation to cater to diverse languages, benefiting the whole humanity.
    \item \textbf{Versatility}: The prompt tuning framework enables a broad spectrum of speech generation tasks, including speech translation, speech inpainting, speech enhancement, etc., to be performed.
    \item \textbf{Efficiency}: The proposed framework offers a unified approach for all speech generation tasks, simplifying the design of downstream models and loss functions for each task. Furthermore, it requires only a small number of trainable parameters, resulting in enhanced computational efficiency. This unified and parameter-efficient framework streamlines the development process for various speech generation tasks.
    \item \textbf{Transferability}: The unified framework is easy to follow and transferable to future advanced speech LMs. This feature holds great promise for efficiency and effectiveness, especially considering the imminent arrival of advanced speech LMs, which will significantly enhance the capabilities of the framework.
    \item \textbf{Affordability}: Our framework is designed to be highly efficient, requiring the training of only 10 million parameters instead of the entire speech language model. This significantly reduces the computational burden and allows the training process to be executed on a GTX 2080 GPU, which is accessible and affordable for universities.
\end{itemize}

The paper is structured as follows: Section 2 introduces related works followed by the proposed speech prompt framework in Section 3. In Section 4, the task descriptions are presented. Section 5 discusses the experimental results and provides an analysis. The limitations and future directions are discussed in Section 6. Finally, Section 7 concludes the paper.
We will open-source the code on the project website \footnote{https://ga642381.github.io/SpeechPrompt/speechgen}.

\section{Related Works}

\subsection{Speech Language Models}

Self-supervised speech representation learning \citep{mohamed2022self} has brought about a revolution in the field of speech processing by facilitating unsupervised learning from large volumes of unlabeled speech data.
By utilizing these self-supervised speech features, it has demonstrated remarkable performance across a wide range of downstream speech processing tasks \citep{DBLP:conf/interspeech/YangCCLLLLSCLHT21, DBLP:conf/acl/TsaiCHHLYDLLSCH22, DBLP:conf/slt/FengDYYLSCHWCWMLL22, wang2023minisuperb}.
Furthermore, by discretizing these self-supervised speech representations, it becomes possible to generate high-quality speech tokens directly from continuous speech. This advancement creates new possibilities in speech processing, enabling the utilization of discrete speech tokens for spoken language modeling~\citep{lakhotia2021generative}.

Speech language models (Speech LMs) leveraging speech tokens have demonstrated exceptional capabilities. Prominent examples include Generative Spoken Language Model (GSLM)~\citep{lakhotia2021generative}, prosody-aware Generative Spoken Language Model (pGSLM)~\citep{DBLP:conf/acl/KharitonovLPACL22}, and Unit mBART \citep{popuri2022enhanced}.
These models have gained significant attention in the field, signaling the forthcoming advancement of speech LMs.
GSLM specializes in performing generative language modeling on discrete units that encode phonetic information extracted by self-supervised speech models such as CPC \citep{oord2018representation}, wav2vec2 \citep{baevski2020wav2vec}, and HuBERT \citep{DBLP:journals/taslp/HsuBTLSM21}. 
Building upon GSLM, pGSLM enhances the model's capabilities by incorporating prosody information. 
Additionally, Unit mBART extends the scope of spoken language modeling to encompass multi-lingual scenarios, facilitating the generation of speech across different languages.
It is worth noting that these models are pre-trained on speech tokens without any text supervision. We refer to these models as \textbf{textless speech LMs}. 

In addition to the textless speech LMs, there exist models that try to mix text and speech modality in a unified model for various speech processing tasks \citep{DBLP:journals/corr/abs-2105-03070, DBLP:conf/acl/AoWZ0RW0KLZWQ0W22, wang2023viola}.
Furthermore, some works incorporate text information to enhance the generation's expressiveness. Models like VALL-E \citep{DBLP:journals/corr/abs-2301-02111, DBLP:journals/corr/abs-2303-03926} and Spear-TTS \citep{DBLP:journals/corr/abs-2302-03540} perform text-to-speech (TTS) in a generative language modeling manner based on neural codes, such as codes from SoundStream \citep{DBLP:journals/taslp/ZeghidourLOST22} and EnCodec \citep{DBLP:journals/corr/abs-2210-13438}.

In this paper, our primary focus lies on textless speech LMs that conduct pre-training without any text supervision.
The reason for this emphasis stems from the fact that textless speech generation is of great significance, given the challenges posed by acquiring labeled text-speech pairs and the absence of text in certain languages \citep{DBLP:journals/corr/abs-2211-06474}.

\subsection{Prompting and Model Reprogramming}
Since the introduction of GPT-3 \citep{brown2020language}, the prompting paradigm \citep{liu2021pre} has emerged as a widely adopted technique for leveraging large language models (LLM) \citep{floridi2020gpt, wei2022chain, zhou2022large, white2023prompt}, or so-called foundation models \citep{DBLP:journals/corr/abs-2108-07258}. Prompting involves utilizing task-specific templates or parameters to guide the language model to perform various tasks without modifying the pre-trained LLM's parameters.

Researchers have explored natural language prompts, often referred to as ``hard prompts,'' \citep{DBLP:conf/emnlp/PetroniRRLBWM19, DBLP:journals/tacl/JiangXAN20, shin2020autoprompt, DBLP:conf/chi/ReynoldsM21} and embedding-based prompts, known as ``soft prompts.'' \citep{DBLP:conf/acl/LiL20, liu2021p, liu2021gpt, DBLP:conf/interspeech/ChangT0L22} In SpeechGen, we adopted soft prompts as the prompting method, for it offers more capability to steer the backbone speech LM \citep{liu2021pre}.
Unlike traditional approaches for model tuning, prompting utilizes a unified feedforward process, allowing a language model to handle multiple tasks simultaneously within the same batch \citep{DBLP:conf/emnlp/LesterAC21}. This unique characteristic offers significant efficiency when deploying models \citep{DBLP:conf/icml/SunSQHQ22}.

On the other hand, similar to prompting, model reprogramming \citep{DBLP:conf/iclr/ElsayedGS19, DBLP:conf/icml/TsaiCH20, DBLP:conf/icml/YangTC21, DBLP:journals/corr/abs-2202-10629} repurposes a pre-trained model by learning a transformation function for its input, thereby enabling the pre-trained model to perform a specific target task. It has been shown reprogramming can be used in speech and audio processing fields, such as spoken command classification \citep{yen2021neural}, music genre classification \citep{hung2023low}, cross-lingual speech recognition \citep{yang2023english}, and dialect identification \citep{radhakrishnan2023parameter}

\subsection{Prompting Speech Models}
SpeechPrompt \citep{DBLP:conf/interspeech/ChangT0L22} was a pioneering work that introduced prompt tuning based on GSLM, enabling it to excel in speech classification tasks such as keyword spotting and intent classification, as well as speech decoding tasks like ASR and slot filling. 
It demonstrated the effectiveness of prompting a speech LM to successfully handle a wide range of speech processing tasks.
Furthermore, in SpeechPrompt v2 \citep{DBLP:journals/corr/abs-2303-00733}, the transferability of the prompting technique is investigated across different speech LMs, specifically from GSLM to pGSLM. 
The study examines the effectiveness of prompt tuning in facilitating a broader range of speech classification tasks, including those involving multilingual and prosody-related aspects.

On the other hand, WAVPROMPT \citep{DBLP:conf/interspeech/GaoNQZCH22} consists of a text LM, GPT-2, and an audio encoder, wav2vec 2.0. The text LM is prompted with audio embeddings and text questions to perform few-shot audio and speech understanding tasks. WAVPROMPT is also a pioneer in studying the prompting paradigm in speech processing. In contrast to SpeechPrompt, which uses a speech LM as its backbone LM, WAVPROMPT employs a text LM as its backbone LM.

Besides prompting speech generative language models, \citep{DBLP:journals/corr/abs-2305-11095} prompts Whisper \citep{radford2022robust}, a supervised speech discriminative model. Whisper is trained on a large amount of paired speech and text data in a multi-task manner. By manually designing task-specific prompts, Whisper demonstrates its zero-shot ability on novel tasks, including audio-visual speech recognition and code-switched speech recognition.
However, the effectiveness of prompts in eliciting generation tasks from speech LMs still remains an open question.

\section{Methodology}
\label{sec:method}
\begin{figure}[h!]
    \centering
    \includegraphics[width=0.98\columnwidth]{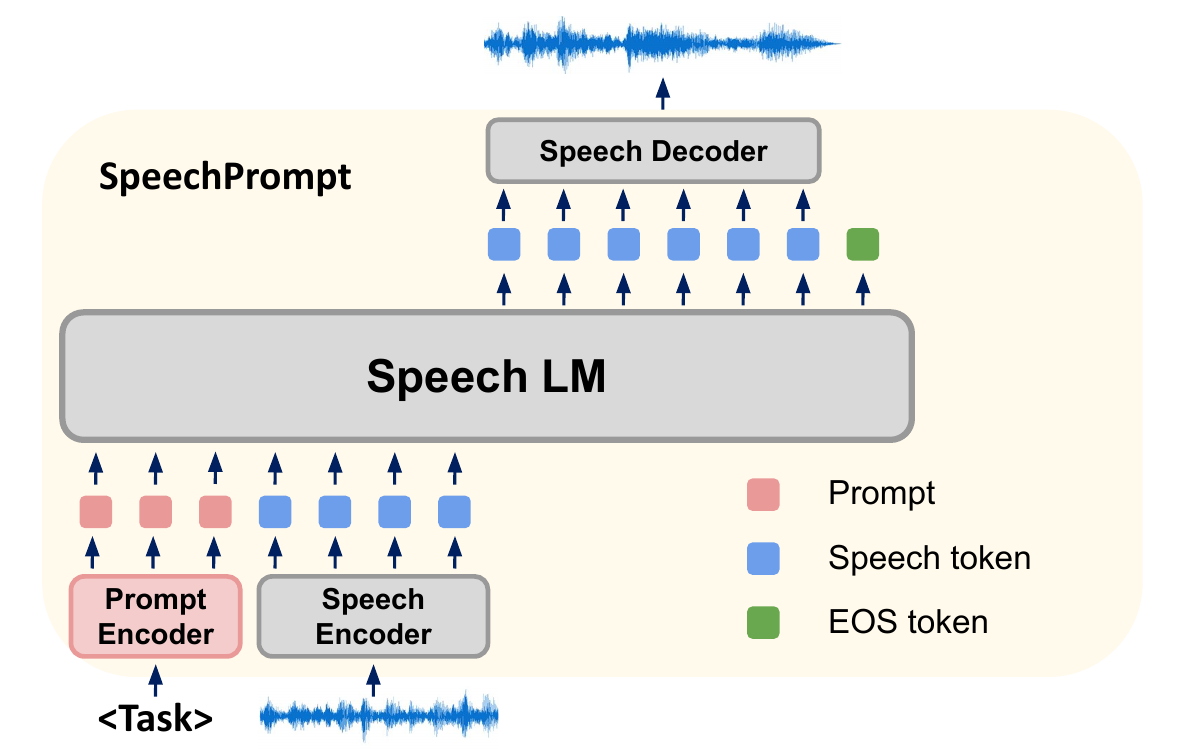}
    \caption{Overview of the SpeechGen framework. SpeechGen utilizes the SpeechPrompt method for generating speech. SpeechGen further includes a speech decoder to generate waveforms, making it versatile for performing various speech-to-speech tasks in a textless manner.}
    \label{fig:framework}
\end{figure}

Figure.\ref{fig:framework} illustrates the entire framework. We introduce SpeechGen, a novel framework that prompts a speech LM for various downstream speech generation tasks. Our method focuses on learning task-specific prompt vectors while keeping the speech LM's parameter fixed. The speech LMs generate the required output for the specific speech generation task effectively by conditioning on both the prompt vectors and input units. The resultant discrete unit outputs are then fed into the unit-based vocoder, which generates corresponding waveforms.

In this section, we delve into the detailed implementation of SpeechGen, focusing on the application of universal prompting techniques to speech generation tasks.

\subsection{Framework}
We implement a direct speech-to-speech framework with discrete units, SpeechGen, as shown in Figure~\ref{fig:framework}, comprised of three elements: a speech encoder, a speech LM, and a speech decoder. Initially, the speech encoder takes the waveform $\bm{s} \in \mathbb{R}^T$ as input and converts it into a unit sequence $\bm{x} = (x_1, \cdots, x_{T’})$, derived from a finite vocabulary. To shorten the sequence length, repeated consecutive units are removed to yield a condensed unit sequence. Next, the speech LM operates as a language model for unit sequences, optimizing the likelihood by predicting the succeeding unit $y_t$ using the preceding units $y_{<t}$ and the unit sequence $\bm{x}$. We use prompt tuning on the speech LM to guide it toward generating the appropriate units based on the task. Finally, the tokens generated by the speech LM are processed by a speech decoder, converting them back into the waveform.

\subsection{Prompt Tuning}
In the approach of prompt tuning, prompt vectors are inserted at the start of the input sequence $\bm{x}$, serving as a guide for the speech LMs during the generation process. The quantity of appended prompts relies on the architecture of the speech LM. Particularly, in a sequence-to-sequence model, prompts are added to the inputs of both the encoder and decoder. In contrast, for an encoder-only or decoder-only architecture, only a single prompt sequence is appended to the input sequence.

For a sequence-to-sequence speech LM (e.g., mBART \citep{popuri2022enhanced}), our technique incorporates a self-supervised learning model (e.g., HuBERT \citep{DBLP:journals/taslp/HsuBTLSM21}) to process both the input and target speech, producing discrete units for the input $\bm{x}$, and the target $\bm{y}$. To form the input sequence $\bm{z}$, we append prompt vectors to the inputs of both the encoder and decoder. We have $\bm{z} = [\bm{p}^E, \bm{x}; \bm{p}^D, \bm{y}]$, where $\bm{p}^E$ is the encoder's prompt vector, and $\bm{p}^D$ is the decoder's prompt vector. The length of each prompt vector is defined by $L \times \text{dim}(h)$, where $L$ is the prompt vector's length, and $\text{dim}(h)$ refers to the dimensionality of the other input discrete units. To boost the guiding capacity of the prompts, we integrate deep prompt tuning \citep{DBLP:conf/acl/LiL20, DBLP:journals/corr/abs-2303-00733} into our method. Specifically, the initial $L$ vectors of the key $K$ and value $V$ are replaced with prompt vectors $\bm{p}^K$ and $\bm{p}^V$ respectively. The updated key and value are calculated as follows:
\begin{align}
K &= \text{Concat}(\bm{p}^K,\bm{x}^j_{l+1:T})\bm{W}^K, \\
V &= \text{Concat}(\bm{p}^V,\bm{x}^j_{l+1:T})\bm{W}^V,
\end{align}
Here, $\bm{x}^j$ denotes the original input of the $j$-th transformer layer, while $\bm{W}^K$ and $\bm{W}^V$ represent the projection matrices utilized in the multi-head attention mechanism. This substitution strategy guarantees that the prompts strongly influence the model's attention mechanism, effectively directing the generation process at multiple layers.

The cross-entropy loss serves as the objective function for all generation tasks, computed by contrasting the model's predictions with the corresponding target discrete unit labels. Importantly, the only trainable parameters in the model are the prompt vectors; the speech LM's parameters remain constant during training, guaranteeing uniform behavior. The inclusion of prompt vectors allows the speech LM to extract task-specific data from the input and boost the likelihood of producing the desired output for the specific speech generation task. This technique allows for the fine-tuning and adaptation of the speech LM's behavior without altering its fundamental parameters.

\vspace{-0.4em}
\section{Task Description}
\begin{figure}[h!]
    \centering
    \includegraphics[width=0.8\columnwidth]{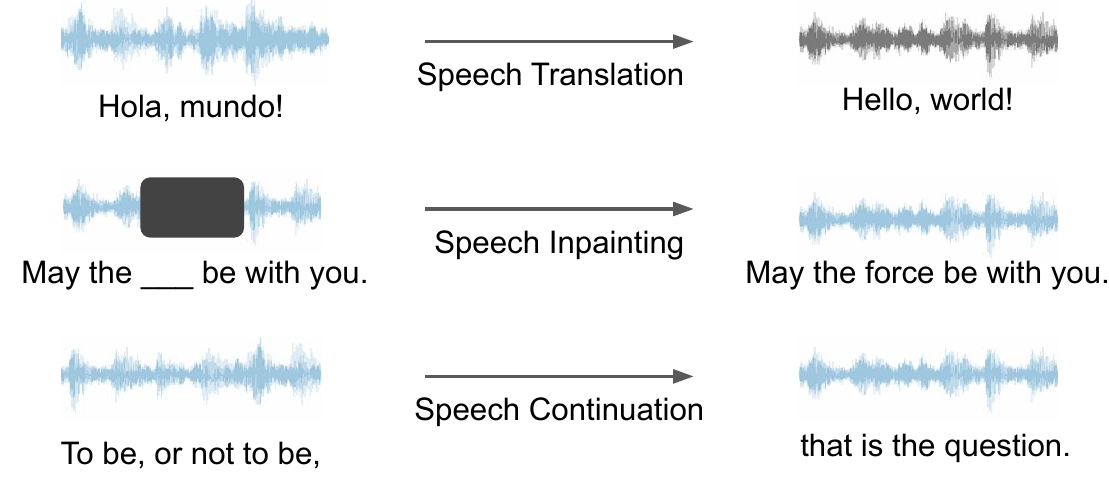}
    \caption{Illustration of three speech generation tasks. During training, there is no text supervision. The shown texts are for demonstration.}
    \label{fig: tasks}
\end{figure}

To showcase the capabilities of our unified framework for speech generation, we present three case studies encompassing speech translation, speech inpainting, and speech continuation tasks. The illustration of the three speech generation tasks is shown in Figure~\ref{fig: tasks}.

\label{sec:task}

\subsection{Speech Translation}

Speech translation (ST) is the process of converting acoustic speech signals from the source language into corresponding speech signals in the target language, enabling communication between individuals who speak different languages.

For our experiments, we utilize the CoVoST2 \citep{wang2020covost} Es-En dataset, which is licensed under CC0. This dataset consists of parallel text data for Spanish (Es) and English (En) translations. 
However, the Spanish portion of the dataset only includes textual data. 
To address this, we employ text-to-speech synthesis techniques in ESPNet \footnote{https://huggingface.co/espnet/kan-bayashi\_ljspeech\_vits} to generate the corresponding Spanish speech for the provided text samples.

\subsection{Speech Inpainting}
\label{ssec:SI}
Speech inpainting is a technique that seeks to restore or fill in missing sections of audio signals. This entails the reconstruction of the damaged portion of the speech recording to produce a seamless and coherent output.

For our experimental setup, we utilized the LibriSpeech dataset \cite{panayotov2015librispeech}. We particularly choose audio segments that surpass a duration of 2.0 seconds as our target speech for the following steps. Using the train-clean-100 subset of LibriSpeech, we generated our training set. The dev-clean subset was employed for the validation set, and the test-clean subset for the test set. Within the selected segments, we adopt a random selection approach to identify a speech segment with a duration that varies between 0.4 and 0.8 seconds. This specific segment is subsequently masked to emulate the absent or corrupted portion in the speech inpainting challenge. As a result, from the train, validation, and test files, we've created 27949, 600, and 600 samples respectively, summing up to 100.17 hours, 1.33 hours, and 1.38 hours in total.




\subsection{Speech Continuation}
Speech continuation (SC) refers to the task of creating a consistent and coherent continuation of a given speech input while preserving the semantic context.

In our SC experiment, we adopted the LJSpeech dataset \citep{ljspeech17}, which contains approximately 24 hours of English speech from a single speaker. The dataset's unique single-speaker attribute provides us with the ability to maintain speaker consistency throughout the speech continuation process.
We divided the LJSpeech dataset into training, validation, and testing subsets. 
Within these subsets, we designated the initial $r$ fraction of each utterance as the seed segment for the speech continuation tasks. We refer to the value of $r$ as the \emph{conditional ratio}.
Our model's objective is to generate an appropriate continuation of the speech, given this seed segment.

\section{Experiments}
\label{sec:expt}
In our experiments, we conduct a case study by using Unit mBART \citep{popuri2022enhanced} as the backbone speech LM in our framework, for Unit mBART is the current open-sourced and state-of-the-art speech LM trained on a large corpus.
It is important to note that the SpeechGen framework is not restricted to a single speech LM. It can be applied to other advanced speech LMs, such as AudioLM \citep{DBLP:journals/corr/abs-2209-03143}, TWIST \citep{hassid2023textually}, and SPECTRON \citep{nachmani2023lms}, which offer additional sophistication but are not yet open-sourced.
With the continuous emergence of further speech LMs in the future, our framework has the potential to be further explored and leveraged to a greater extent.

\subsection{Experimental Setup}
In the implementation of SpeechGen, we employ Unit mBART as the underlying backbone SLM, offering a robust foundation with impressive performance.
We apply \emph{deep prompt tuning} \citep{DBLP:conf/acl/LiL20, liu2021p, DBLP:conf/interspeech/ChangT0L22}, where the prompts are prepended at each transformer layer's input. The prompts are applied to both the encoder and decoder of the Unit mBART. 
The prompt length is set to be 200 for both the encoder and decoder, corresponding to about 10M trainable parameters in total.
In order to highlight the convenience and flexibility of the prompting paradigm and streamline the entire pipeline, we do not conduct a hyperparameter search.
Additionally, we adopt the pre-trained unit-based HiFi-GAN neural vocoder \citep{popuri2022enhanced} to generate high-quality waveforms. 

\subsection{Experimental Results and Analysis}

\subsubsection{Speech Translation}
In this section, we show the speech translation results from Spanish to English using SpeechGen.
We adopt sacrebleu \footnote{https://github.com/mjpost/sacrebleu} to calculate the BLEU scores and the BLEU scores are shown in Table~\ref{tab:blue scores}.
To provide a more tangible demonstration of the translation performance, we present several demos in Table~\ref{tab:translations}.
The showcased demos demonstrate that the model predictions capture the core meanings of the ground truth.

\begin{table}[htbp]
  \centering
  \caption{BLEU Scores}
  \label{tab:blue scores}
  \begin{tabular}{cccc}
    \toprule[1pt]
    \textbf{BLEU-1} & \textbf{BLEU-2} & \textbf{BLEU-3} & \textbf{BLEU-4} \\
    \hline
    43.8 & 30.4 & 21.8 & 15.9 \\
    \toprule[1pt]
  \end{tabular}
\end{table}

\begin{table}[htbp]
  \centering
  \caption{Examples for translation from Spanish to English}
  \label{tab:translations}
  \begin{tabular}{p{6.5cm} p{6.5cm}}
    \toprule[1pt]
    \textbf{Ground truth} & \textbf{Model prediction} \\
    \hline
    \\
    The origin of the name of the county is uncertain. & Origin of the name of the county is uncertain. \\
    \\
    \hline
    \\
    Lastly, the play will devote a reflection to the relationship between art and rebellion. & And lastly the work will devote a reflection to the relationship between art and rebellion. \\
    \\
    \hline
    \\
    It is around thirty kilometers away from the regional capital city. & Just one hundred forty kilometers from the regional capital. \\
    \\
     \hline
     \\
    They were easily recognized by the use of the armor and the "Farina" helmet. & They were frequently recognized for the use of armor and the cascade. \\
    \\
    \hline
    \\
    They played in cover bands but decided to create their own music. & They played in mandates but they decided to create their own music. \\
    \\
    \toprule[1pt]
  \end{tabular}
\end{table}

\begin{figure}[ht!]
    \centering
    \includegraphics[width=0.98\columnwidth]{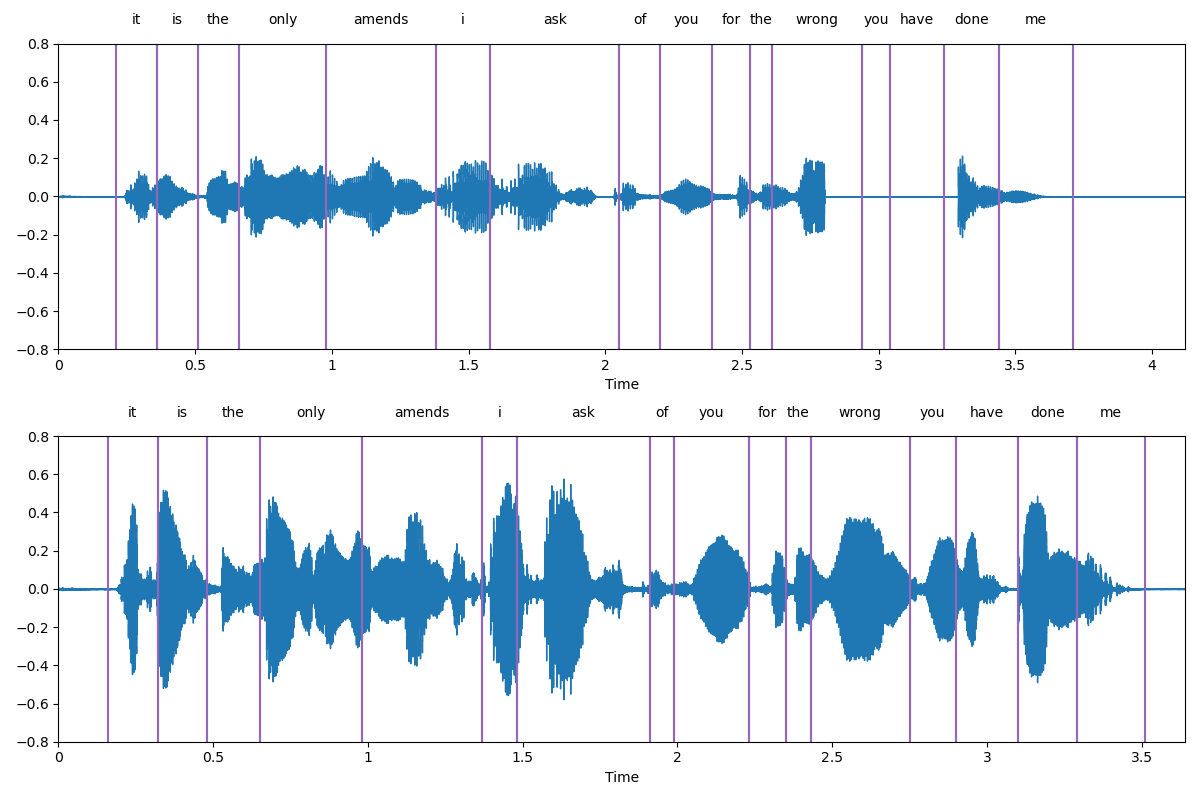}
    \caption{The figure comprises two graphs: the upper plot represents the corrupted waveform, while the lower one depicts the waveform as generated by SpeechGen. Word boundaries are depicted within the plots and the transcriptions, derived from the ASR, are presented above each plot. For the corrupted waveform, its transcription and word boundaries are generated using the original waveform.}
    \label{fig: inpainting}
\end{figure}
\subsubsection{Speech Inpainting}
In our research, we evaluated the performance of our SpeechGen model within the context of the speech inpainting task, focusing on the test set of the dataset detailed in section \ref{ssec:SI}. Our objective was to ensure a fair assessment across both real and synthesized data. Consequently, rather than relying on a pretrained model which had only been trained on real data, we opted to train an ASR (Automatic Speech Recognition) model. For the ASR training, we employed the S3PRL toolkit\footnote{https://github.com/s3prl/s3prl} and adhered to its default configuration. Furthermore, we utilized the FBANK upstream model for our ASR.

To quantify the extent of successful restoration of corrupted speech segments, we employed two metrics: word error rate (WER) and character error rate (CER). Our results revealed that the SpeechGen model yielded a 25.42\% WER and a 13.85\% CER as shown in Table \ref{tab:inpainting}. Interestingly, these metrics closely mirrored those of the corrupted speech, which registered at 27.96\% WER and 13.47\% CER. This indicates that our model's performance in speech inpainting is not optimal.

Nevertheless, there are instances, as depicted in Fig. \ref{fig: inpainting}, where our model adeptly restored accurate words in the masked sections of original speech. Yet, the overall outcomes have not lived up to our aspirations, signaling the need for further refinement in our methodology.\footnote{We would like to acknowledge the discrepancies observed in our previous experiment which led to unintended results in an earlier version of this work. We sincerely apologize for this oversight and are dedicated to rectifying the issue.}

\begin{table}[h!]
    \centering
    \caption{Performance evaluation of SpeechGen on the speech inpainting task}
    \begin{tabular}{cccll}
    \toprule[1pt]
    \multicolumn{1}{l}{} & WER     & CER      \\
    \hline
    SpeechGen         & 25.42\% & 13.85\%  \\
    Corrupted            & 27.96\% & 13.47\%  \\
    Original                & 18.02\% & 5.48\%   \\
    \toprule[1pt]
    \label{tab:inpainting}
    \end{tabular}
\end{table}

\subsubsection{Speech Continuation}
We measure the speech continuation performance in two objective metrics, perplexity (PPX) and auto-BLEU \citep{lakhotia2021generative}. Before evaluation, the speech will first be transcribed into text with an off-the-shelf ASR model\footnote{~``Wav2vec2\_large\_lv60k'' with CTC decoder available on PyTorch}. Perplexity is calculated by a pre-trained language model \footnote{~``transformer\_lm.wmt19.en'' available on Fairseq}, and auto-BLEU is calculated within a sentence, and it indicates the \textbf{diversity} of a generated speech.

For SpeechGen, we tested the speech continuation with a different conditioned ratio, $r$. The conditioned ratio means the ratio of the seed segment in each utterance. For example, when $r = 0.5$, it means that the first half of the utterance will be conditioned by SpeechGen, and SpeechGen will keep generating novel speech (red text in Table \ref{sc-example}).

Table~\ref{sc-metric} shows the result. We can observe that the diversity keeps comparable across different scenarios, indicating that SpeechGen can generate novel speech \emph{as diverse as the original speech}, rather than the stuttering speech presented in GSLM \citep{lakhotia2021generative}. 
Furthermore, by observing the examples in Table~\ref{sc-example}, we can see that the generated novel speech is basically grammar correct and semantic related to the seed segment.

\begin{table}[htbp]
    \centering
    \caption{Evaluation of SpeechGen for the speech continuation task. Original: The original utterances in the dataset}
    \begin{tabular}{ccccc}
    \toprule[1pt]
    \multicolumn{1}{l}{} & PPX     & auto-BLEU-1 & auto-BLEU-2 & auto-BLEU-3       \\
    \hline
    Original             & 202.92   &0.939  &0.648  &0.391 \\
    SpeechGen ($r = 0.25$) & 543.8	&0.937	&0.658	&0.405 \\
    SpeechGen ($r = 0.5$)  & 420.49	&0.938	&0.647	&0.388 \\
    SpeechGen ($r = 0.75$) &283.03	&0.936	&0.643	&0.396 \\
    \toprule[1pt]
    \label{sc-metric}
    \end{tabular}
\end{table}
\begin{table}[h!]
  \centering
  \caption{Examples of speech continuation with different conditional ratios. The \textcolor{red}{text in red} is the speech generated by SpeechGen conditioned on the seed segment, which is text in black. The provided texts are the recognition results for reference. SpeechGen performs speech continuation in a speech-to-speech manner.}
  \begin{tabular}{p{2cm} p{11cm}}
    \toprule[1pt]
    \textbf{Cond. Ratio} & \textbf{Speech Continuation Result} \\
    \hline
    \\
    0.25 &  He assassinated the president \textcolor{red}{and gave mister johnson the last charge of improvement in his writing possible three point eight nine.} \\
    \\
    \hline
    \\
    0.25 &  If anybody really wanted \textcolor{red}{to interfere once for some time even if he got anything back to the other folk} \\
    \\
    \hline
    \\
    0.25 & Under standard procedures \textcolor{red}{the receivers have inspected trade here by this time have recently developed they fully recognize did their business.}\\

    \\
    \hline
    \\
    0.5 &  Who had been greatly upset by her experience was able to view a \textcolor{red}{longer time the great farce and receive good care of the mutter.} \\
    \\
    \hline
    \\
    0.5 &  Childless parents widows and helpless orphans broken and controlled \textcolor{red}{by the master and sentence pursuit life apt to paradise.} \\
    \\
    \hline
    \\
    0.5 & But these king's witnesses were also put at times into the press yard \textcolor{red}{and charged with the service available on a second charge to them.}\\
    \\
    \hline
    \\
    0.75 &  And the obvious bulk of the package which he intended to bring to work \textcolor{red}{was confirmed} \\
    \\
    \hline
    \\
    0.75 &  Then they set to building and began by bricking the borders of the moat after which they proceeded \textcolor{red}{to our own places} \\
    \\
    \hline
    \\
    0.75 & Still watching and waiting for the first chance they ceased when the clerks \textcolor{red}{left the office}\\
    \\
    \toprule[1pt]
    \label{sc-example}
  \end{tabular}
\end{table}

\section{Limitations and Future Directions}
\label{sec:limitations}
Here we list some limitations and future directions:
\begin{itemize}
    \item \textbf{Develop advanced speech language models}: Speech language models are currently in their nascent stage of development compared to text-based language models. The proposed prompt framework, although effective in motivating speech language models, may not achieve exceptional performance. However, with advancements in speech language models, such as the transition from GSLM to Unit mBART, there has been a significant improvement in prompt performance. Particularly, tasks that were previously challenging for GSLM now exhibit improved performance with Unit mBART. We anticipate the emergence of even more promising speech language models in the future.
    \item \textbf{Beyond content information}: The current speech language models do not fully capture speaker and emotion information, which poses a challenge for the current speech prompt framework in effectively handling such information. To address this limitation, we plan to explore the integration of plug-and-play modules that specifically incorporate speaker and emotion information into the framework. Looking ahead, we anticipate that future speech language models will incorporate and leverage these additional factors to enhance their performance and better handle speaker and emotion-related aspects in speech generation tasks. Google's latest speech LM \citep{nachmani2023lms} tries to include such information.
    \item \textbf{Possibility of prompt generation}: For prompt generation, we have the flexibility to integrate various types of instructions, including text and image instructions, as potential choices. One approach is to train a network using images or texts as inputs, instead of the trainable prompts in this paper. This trained network can then serve as a prompt generator, enhancing the versatility of our framework.
    \item \textbf{Hyperparameter search}: The optimization of hyperparameters can also contribute to improved performance. In our preliminary research, we have not yet conducted an extensive search for hyperparameters, such as prompt length and batch size. However, in future investigations, we aim to enhance performance by refining these hyperparameters to achieve better results.
\end{itemize}

\section{Conclusion}
\label{sec:conclusion}
In this paper, we investigate the utilization of prompt tuning to facilitate the performance of speech language models (LMs) across a wide range of speech generation tasks. 
Our approach is implemented within a unified framework called SpeechGen, which is with around 10M trainable parameters.
The proposed framework exhibits several desirable properties, including its textless nature, versatility, efficiency, transferability, and affordability.
To demonstrate the capabilities of our framework, we take Unit mBART as a case study, and conduct experiments on three distinct speech generation tasks: speech translation, speech inpainting, and speech continuation.
We also discuss the limitations and future directions of prompting speech LMs.
With the imminent advent of advanced speech LMs, our unified framework holds immense potential in terms of efficiency and effectiveness by standing on the shoulders of giants.

\bibliographystyle{elsarticle-harv}
\bibliography{reference}




\end{document}